\documentclass[twocolumn, numberedappendix, twocolappendix, apj]{openjournal}
\usepackage{newtxtext,newtxmath}
\usepackage{amsmath}
\usepackage{amssymb}
\usepackage{graphicx}
\usepackage{xspace}
\usepackage[table]{xcolor}
\usepackage[colorlinks=true, allcolors=blue]{hyperref}
\usepackage{fontawesome5}
\usepackage{multirow}
\usepackage{listings}
\usepackage{xcolor}
\usepackage[normalem]{ulem}

\definecolor{codegreen}{rgb}{0,0.6,0}
\definecolor{codegray}{rgb}{0.5,0.5,0.5}
\definecolor{codepurple}{rgb}{0.58,0,0.82}
\definecolor{backcolour}{rgb}{0.95,0.95,0.92}
\lstdefinestyle{mystyle}{
    backgroundcolor=\color{backcolour},   
    commentstyle=\color{codegreen},
    keywordstyle=\color{magenta},
    numberstyle=\tiny\color{codegray},
    stringstyle=\color{codepurple},
    basicstyle=\ttfamily\footnotesize,
    breakatwhitespace=false,         
    breaklines=true,                 
    captionpos=b,                    
    keepspaces=true,                 
    numbers=left,                    
    numbersep=5pt,                  
    showspaces=false,                
    showstringspaces=false,
    showtabs=false,                  
    tabsize=2
}

\lstdefinelanguage{Julia}%
  {morekeywords={abstract,break,case,catch,const,continue,do,else,elseif,%
      end,export,false,for,function,immutable,import,importall,if,in,%
      macro,module,otherwise,quote,return,switch,true,try,type,typealias,%
      using,while},%
   sensitive=true,%
   morecomment=[l]\#,%
   morecomment=[n]{\#=}{=\#},%
   morestring=[s]{"}{"},%
   morestring=[m]{'}{'},%
}[keywords,comments,strings]%

\begin{document}

\title{Impact of lensing of gravitational waves on the observed distribution of neutron star masses}

\author{
Sofia Canevarolo,$^{*,1}$
Loek van Vonderen,$^{1}$
Nora Elisa Chisari$^{1}$
}
\email{$^*$s.canevarolo@uu.nl}
\affiliation{$^{1}$Institute for Theoretical Physics, Utrecht University, Princetonplein 5, 3584 CC, Utrecht, The Netherlands.}
\date{\today}

\begin{abstract}

The distribution of masses of neutron stars, particularly the maximum mass value, is considered a probe of their formation, evolution and internal physics (i.e., equation of state). 
This mass distribution could in principle be inferred from the detection of gravitational waves from binary neutron star mergers. Using mock catalogues of $10^5$ dark sirens events, expected to be detected by Einstein Telescope over an operational period of $\sim1\, \rm year$ , we show how the biased luminosity distance measurement induced by gravitational lensing affects the inferred redshift and mass of the merger. This results in higher observed masses than expected. Up to $2\%$ of the events are predicted to fall above the maximum allowed neutron star mass depending on the intrinsic mass distribution and signal-to-noise ratio threshold adopted. The underlying true mass distribution and maximum mass could still be approximately recovered in the case of bright standard sirens.\\
\end{abstract}


\maketitle



\section{Introduction}

Knowledge about the neutron star (NS) mass distribution is of crucial importance to answer important open questions regarding the NS physics. The shape of the NS mass distribution is closely related to the different formation and evolution mechanisms of such objects \citep{Schwab2010, Valentim2011,Pejcha2012,Kiziltan2013}. Moreover, the existence of a maximum mass for NSs was first theoretically derived by \citet{Tolman1939,Oppenheimer1939} by
solving the Tolman–Oppenheimer–Volkoff (TOV) equation, and subsequently refined by \citet{Rhoades1974, Kalogera1996}. The exact value of the maximum mass is still under debate, as it strongly depends on the Equation of State (EoS) of NSs, which is unknown \citep[see reviews][]{Burgio2021,Koehn2024}. 

Besides invaluable information on the subatomic matter in the NS interior, the maximum mass of NSs might establish a dividing line between the NS and the black hole (BH) populations. Analyses of current gravitational wave (GW) events distinguish between BHs and NSs by comparing the mass of the two components of the binary with a reference value for the maximum mass of NSs. This procedure is supported by the expectation of a lower mass gap between the heaviest NSs and the least  massive astrophysical BHs, with the exclusion of those formed from an earlier generation of mergers \citep{Bailyn1998,Ozel2010,Farr2011, Kreidberg2012, Liu2021, Gayathri2023}. The upper bound of the lower mass gap is the most uncertain and mainly considered an observational fact. Possible explanations rely on the physics related to the supernovae explosion mechanisms. \citet{Belczynski2012} relates the presence of the mass gap to the growth timescale of the instabilities driving the supernovae explosion, finding hints that a rapid explosion model is preferred to explain the gap. However, observational bias may also provide an explanation to the gap \citep{Kreidberg2012, deSa2022}. Therefore the existence of this gap is still under debate, especially given a few observations, most notably the event GW190814 \citep{Abbott2020} and the very recent event GW230529 \citep{LigoVirgo2024}, both with a mass of one of the components likely to fall into the gap. 

The first three observing runs (O1, O2 and O3) of LIGO and Virgo GWs observatories recorded a small number of events in which at least one of the two objects in the binary system was likely to be a neutron star. The Gravitational Wave Transient
Catalog 3 (GWTC-3) \citep{Abbott2023} contains two binary neutron stars (BNS) and two neutron star-black hole (BNBH) events. Despite the paucity of NS observations, these events provided a new way to learn about matter in extreme conditions and NS physics, complementary to the already on-going observations in the electromagnetic (EM) spectrum \citep{Bogdanov2019,Agazie2023,Shamohammadi2023}.

As the current GW detectors are entering another phase of runs, and the construction of a new, third generation (3G), of ground- and space-based detectors is planned for the next decade, we expect the number of observed GW events involving NSs to increase enormously. The Einstein Telescope (ET) is expected to detect $\sim 10^5$ BNS events per year, up to very high redshifts \citep{Punturo2010, Maggiore2020,Branchesi2023}. These GW events will allow stringent constraints to be placed on the NS mass function and maximum mass. 
 
On the other hand, as more and more distant events are observed, GWs are more likely to have been lensed during their path towards our detectors by the fluctuations of the matter distribution of the Universe \citep{Bartelmann01}. Similarly to the case of light, a few of these GWs can undergo the phenomenon of strong lensing, if the alignment between source-lens-observer is favorable \citep{Liao2022}, or weak lensing in all the other cases \citep{Cutler2009,Camera2013,Congedo2019, Mukherjee2020,Balaudo2023}. Within the geometric optics approximation, the effect of lensing on GWs consists in a frequency-independent modification of their amplitude \citep{Takahashi2003,Laguna2010, Bertacca2018}. With the amplitude being inversely proportional to the luminosity distance of the GW source, this directly translates to a biased estimation of the latter. 

GWs coming from the inspiral and merger of compact objects provide an estimate of the redshifted masses, i.e. the source-frame masses shifted by a redshift factor, of the two components of the binary system \citep{Cutler1993,Finn1993}. Consequently, information about the redshift of the event is needed to access the true masses of the components, which can be used to characterize the astrophysical population of NSs, in particular its mass distribution. In the absence of an independent estimation of the redshift of the source, for example through the detection of an EM counterpart, the biased luminosity distance spoils the estimate of the redshift and, in turn, of the true masses of the components of the binary system. As a consequence, far and light sources might be misinterpreted as close and massive ones. Searches for this effect in already available data are ongoing \citep{Diego2021,Abbott2023l}.

In this paper, we assess the impact of gravitational lensing on the mass distribution of NSs, working with mock catalogues relevant for ET and assuming that the mass-redshift degeneracy cannot be broken by independent observations. Given that we expect the great majority of BNS events observed with current and future GW detectors to satisfy this assumption, it is of crucial importance to understand the impact that lensing can have on the inferred NS mass distribution. Previous works such as \citet{Dai2017}, \citet{Oguri2018} and \citet{He2022} focused on the impact of lensing on the black hole mass distribution, and they found evidence that lensing affects the high-mass tail of the distribution. Moreover, \citet{Graham2023} showed that BNS events can be magnified into the mass gap. Such results motivate a further analysis for NSs in the era of 3G detectors, given the importance of the NS mass distribution and its maximum mass to answer fundamental questions on the physics of the interiors of NS.

We adopt a flat Friedmann–Lemaître–Robertson–Walker (FLRW) cosmology with the following fiducial values of the parameters: $h=H_0/(100\, \rm km \,s^{-1}Mpc^{-1})=0.705$ for the Hubble constant, $\Omega_{\rm m} = 0.274,$ for the matter (baryons plus dark matter) density parameter, $\Omega_{\rm DE}=1-\Omega_{\rm m}$ for the dark energy density parameter and $\sigma_8=0.812$ for the amplitude of the matter fluctuations. These values correspond to those adopted in the simulations of \citet{Takahashi2011}, which we use to model gravitational lensing. 

The paper is organised as follows. In Section~\ref{Sec:Theory}, we outline the impact of gravitational lensing on GW events and in particular on the mass estimation of the progenitors objects in the binary system. In Section~\ref{Sec:Mocks}, we build mock catalogues of lensed BNS events relevant for ET. In Section~\ref{Sec:Results}, we present our results.
In section~\ref{Sec:Discussion} we discuss the relevance of our results and the limitations of our methods. Finally, we conclude in Section~\ref{Sec:Conclusion}.

\begin{figure}
\centering    
\includegraphics[width=0.45\textwidth]{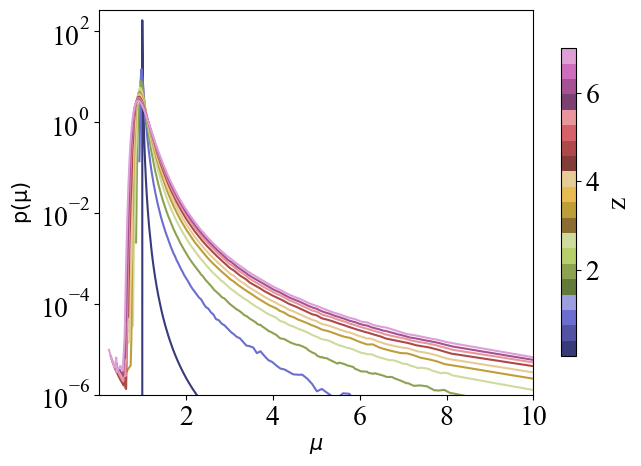}
\caption{Probability distribution function of the lensing magnification for different source redshifts $z$, as obtained from the results of the ray-tracing simulations of \citet{Takahashi2011}, augmented by the lognormal model of \citet{Marra2013} for $z<1$.}
\label{fig:pmu}
\end{figure}

\section{Impact of the lensing biased luminosity distance on the observed masses of dark sirens}
\label{Sec:Theory}

In the GW signal, the redshift of the source $z$ and the source-frame masses of the two components of the binary system $m_{1,2}$ are degenerate, such that only the combination ${m^z_{1,2}=m_{1,2}(1+z)}$ can be constrained from the GW event. The degeneracy is broken if the GW event is accompanied by a measurable EM counterpart, thanks to which an independent estimate of the redshift of the source can be obtained. These ``bright standard sirens'' are of particular interest for their application to cosmology. They allow us to test standard cosmological model through the distance-redshift relation \citep{Holz2005, Dalal2006}. The measurement of the EM counterpart is challenging and it requires synergies with EM observatories, such that only a small fraction of BNS events is expected to be bright standard sirens.

The great majority of GW detections are dark sirens, i.e. not accompanied by any EM counterpart. In this case, the mass-redshift degeneracy can still be broken through the measurement of tidal effects 
if at least one of the two objects is a NS and an EoS is assumed \citep{Messenger2011, Pang2020, Ghosh2022}. However, this requires a high precision measurement of the GW phasing, such that even for ET, only a small fraction of events is expected to provide percent level measurements of the tidal deformability parameter \citep{Branchesi2023}. 

In this work, we consider dark siren events for which the mass-redshift degeneracy cannot be broken by any external information. Working in the geometric optics approximation, which is valid when the wavelength of the GW is much smaller than the size of the lens, the observed luminosity distance is \citep{Kocsis2006}:
\begin{equation}
    d_{\rm L}^{\rm obs}=\frac{d_{\rm L}}{\sqrt{\mu}},
\end{equation}
where $\mu$ is the lensing magnification and $d_{\rm L}$ is the true luminosity distance of the event, i.e. the one that would be measured in a homogeneous universe. In a flat FLRW universe the luminosity distance $d_{\rm L}$ is given by:
\begin{equation}
    d_{\rm L}(z)=\frac{c\, (1+z)}{H_0}\int_0^z\frac{dz'}{\sqrt{\Omega_{\rm m}(1+z')^3+\Omega_{\rm DE}}}, 
    \label{dl-z}
\end{equation}
where $c$ is the speed of light.

The redshift of the event can be estimated by inverting the distance-redshift relation in equation~\eqref{dl-z}, and assuming a cosmology, i.e. $z=z(d_{\rm L})$.
However, neglecting lensing in the analysis, we would infer the redshift of the event from the observed luminosity distance \citep{Cutler1993,Finn1993}:
\begin{equation}
    z^{\rm biased}=z(d_{\rm L}^{\rm obs}),
    \label{zbiased}
\end{equation}
which is biased by lensing. 

Using this latter estimate of the redshift to break the mass-redshift degeneracy, we find that the observed masses are \citep{He2022}:
\begin{equation}
    m_{1,2}^{\rm biased}= \frac{1+z}{1+z^{\rm biased}}\, m_{1,2}:=\mu_{\rm m}\, m_{1,2}.
    \label{observed masses}
\end{equation}
and we call $\mu_{\rm m}$ the mass bias factor to highlight the fact that it can amplify or reduce the value of the biased masses \footnote{This  differs from \citet{He2022} where $\mu_{\rm m}$ is referred to as the amplification factor.} .

In summary, neglecting lensing in the analysis of the GW event, and in absence of an independent estimate of the redshift of the source, the redshift is estimated from the observed distance estimator as given in equation \eqref{zbiased}. This is noisy, biased, and conditioned on a fixed cosmology. Any impact of lensing on the inferred mass distribution is driven by the particular method used to estimate the redshift. In particular, for values of $\mu$ greater than unity, the masses are misinterpreted as heavier than they really are. This is particularly interesting in the case of NSs, for which theory predicts a maximum mass in their intrinsic distribution. 

\section{Mock Neutron Star Population}
\label{Sec:Mocks}

Following \citet{Canevarolo2023}, we build a mock catalogue of lensed BNS events as expected to be detected by ET. We draw a collection of redshifts, $\{z_i\}$, for our events, from the following theoretical distribution:
\begin{equation}
    \begin{split}
        p^{\rm (th)}(z)=\frac{1}{1+z}\frac{4 \pi c \chi^2(z)}{H(z)}&\int_{t_{\rm d}^{\rm MIN}}^{t_{\rm d}^{\rm MAX}} {\rm d}t_{\rm d} \Psi(z_{\rm f}(z,t_{\rm d})) \\
        & \times \bigg( \log\bigg(\frac{t_{\rm d}^{\rm MAX}}{t_{\rm d}^{\rm MIN}}\bigg)t_{\rm d}\bigg)^{-1},
    \end{split}
    \label{pth}
    \end{equation}
where $\chi(z)=c \int_0^z \frac{dz}{H(z)}$ is the comoving distance, ${H(z)=H_0\sqrt{\Omega_{\rm m}(1+z)^3+\Omega_{\rm DE}}}$ is the Hubble parameter, $t_{\rm d}^{\rm MIN}=20\; \rm Myr$ and $t_{\rm d}^{\rm MAX}=10\; \rm Gyr$; $\Psi(z_{\rm f}(z,t_{\rm d}))$ is the Madau-Fragos star formation rate \citep{Madau2017}:
    \begin{equation}
        \Psi(z_{\rm f})=0.01\frac{(1+z_{\rm f})^{2.6}}{1+[(1+z_{\rm f})/3.2]^{6.2}} \;\; \rm M_{\sun} Gpc^{-3} yr^{-1},   
    \end{equation}
and it is integrated over the possible time delays $t_{\rm d}$ between the redshift of the formation $z_{\rm f}$ and merger of the binary system. The probability for the time delay is assumed to be $p(t_{\rm d}) \propto \frac{1}{t_{\rm d}}$ \citep{Regimbau2012,Mukherjee2021, Borhanian2022}. 

A small fraction of the total events is expected to be bright standard sirens, i.e. events accompanied by a measurable EM counterpart. The observation of such multi-messenger events is conditioned by the ability of EM facilities to detect the counterpart, which may come in the form of a Gamma Ray Burst with an afterglow emission or a kilonova, and to identify the host galaxy of the GW source \citep{Belgacem2019,Yu2021,Ronchini2022}. This induces selection effects on the redshift distribution of bright standard sirens that strongly depend on the details of the EM observation. In our work, we simply assume that bright sirens can have a maximum redshift of $z=2$ \citep{Congedo2022}, based on the coverage that future spectroscopic galaxy surveys are expected to reach \citep{Laureijs2011}. 

\begin{figure}
\centering    
\includegraphics[width=0.45\textwidth]{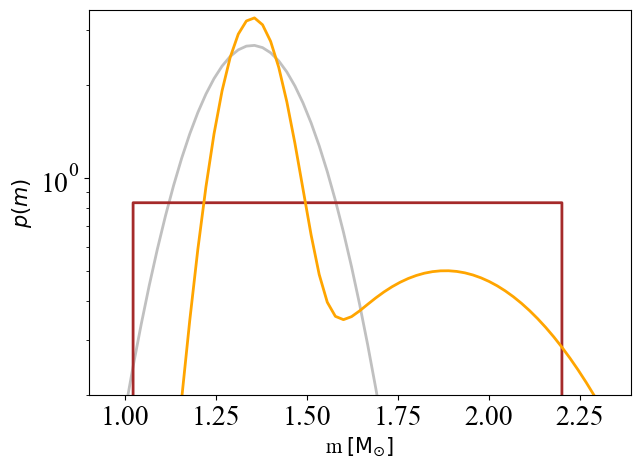}
\caption{Intrinsic neutron star mass distributions considered in this work. The silver line represents a sharply peaked Gaussian distribution, the brown line a uniform distribution and the orange line a double Gaussian. }
\label{fig:pm_distrs}
\end{figure}

For each event, we randomly draw one value of the lensing magnification from its probability distribution function (PDF). We use the lensing PDF provided by \citet{Takahashi2011} via ray-tracing gravity-only simulations. In particular, we use the highest resolution simulations with $1024^3$ number of particles and a grid size of $r_{\rm grid}=3\, {\rm kpc}/h $. This PDF is publicly available\footnote{\label{foot:takahasi/raytracing}\url{http://cosmo.phys.hirosaki-u.ac.jp/takahasi/raytracing/open_data/}} for 7 different source redshifts $z$, spanning from $z=1$ to $z=20$. Below $z=1$, we adopt a log-normal model for the lensing convergence field, $\kappa$, provided by \citet{Marra2013}, together with the cosmology-dependent fitting functions for the moments of the distribution. We fix the moments of the log-normal distribution with the cosmological parameters corresponding to the values used in the ray-tracing simulations of \citet{Takahashi2011}, and we convert the lensing convergence to the lensing magnification using the relation \citep[Figure 8]{Takahashi2011}:
\begin{equation}
    \mu=\frac{1}{(1-\kappa)^2},
    \label{mu-kappa}
\end{equation}
which neglects the effect of the shear. In summary, to construct our lensing magnification PDF, we interpolate between the log-normal model of \citeauthor{Marra2013}, which we assume valid for the regime $z<1$, and the results of the ray-tracing simulations of \citeauthor{Takahashi2011} for $z\geq1$. Our resulting magnification PDFs at different redshifts are shown in Figure~\ref{fig:pmu}. As expected, it peaks at $\mu \sim 1$ reflecting the fact that voids are predominant in the Universe and it has a tail of high magnifications due to the rare overdensities which becomes more relevant with increasing redshift. Finally, the increase of probability for $\mu \lesssim 0.75$ is due to demagnified (Type III) images of strongly lensed events \citep{Takahashi2011}.

To complete our mock catalogue, we also draw the masses of the two components of the BNS. We consider three redshift-independent intrinsic distributions for the NS masses. More details about the mass distributions are given in subsection~\ref{subsec:mass distributions}. We normalise the distributions in the range $[1 \rm M_{\sun};m_{\rm max}]$, and consider different values of maximum NS mass $m_{\rm max}$. We refer to ``mass gap events'' as the events in which only one of the two components of the binary obtains a lensed mass above $m_{\rm max}$. ``Double mass gap events'' are those where both components acquire an observed mass larger than $m_{\rm max}$.

Finally, we draw values for the angles that determine the position of the binary in the sky $(\theta;\phi)$, the inclination of the binary $\iota$, and the polarization angle $\psi$. In particular, we draw $\theta$ and $\iota$ from an isotropic distribution in the interval $[0, \pi]$, and $\psi$ from a uniform distribution in the same range. For the case of the triangular configuration of ET, the detector response function is independent on the angle $\phi$ \citep{Regimbau2012}. For the subgroup of events that we label as bright standard sirens, we restrict $ \iota < 20 \degr$, assuming that the EM counterpart is beamed.

Having a full catalogue of BNS events, we compute the signal-to-noise ratio (SNR) $\rho$ of each event as
    \begin{equation}
        \rho^2=\mu \frac{5}{6} \frac{(G\mathcal{M}(1+z))^{5/3}}{c^3 \pi^{4/3}d_{\rm L}^2(z)}\mathcal{F}^2(\theta,\iota,\psi)\int_{f_{\rm min}}^{f_{\rm max}} df \frac{f^{-7/3}}{S_{\rm n}(f)},
    \label{SNReq}
    \end{equation}
where $S_{\rm n}(f)$ is the one-sided noise power spectral density of the detector; $\mathcal{M}=\frac{(m_1 m_2)^{3/5}}{(m_1+m_2)^{1/5}}$ is the chirp mass of the binary system; $\mathcal{F}^2(\theta,\iota,\psi)$, the function that describes the response of the detector \citep{Regimbau2012}; $G$ is Newton's gravitational constant. 

Assuming a triangular configuration for ET, we obtain the total SNR summing over the three independent detectors \citep{Cutler1994}, considering the same sensitivity curve for each of them: the ET-D noise curve\footnote{\label{foot:ETwebsite}\url{https://www.et-gw.eu}} \citep{Hild2011}. Notice that, to account for possible selection effects to due lensing, we also included the lensing magnification $\mu$ in the calculation of the SNR. Moreover, equation \eqref{SNReq} is an approximation valid only during the inspiral. Therefore, when integrating over frequency, we consider as boundaries $f_{\rm min}=1\,\rm Hz$ and $f_{\rm max}=\frac{1}{6\sqrt{6}(2\pi)}\frac{c^3}{G(1+z)(m_1+m_2)}$, assumed as the frequency at the last stable orbit.

Finally, we keep only the events whose SNR exceeds a threshold set for the detection $\rho_{\rm lim}$, which we assume to be 8 \citep{Regimbau2015}. In this way, we build mock catalogues containing $10^5$ BNS events, which are expected to be observed in $\sim 1\, \rm year$ of observational time of ET. Of the total number of events, we also assume that $300$  multi-messenger events will be detected \citep{Ronchini2022, Branchesi2023}.

\subsection{Neutron star mass distributions}
\label{subsec:mass distributions}

Current knowledge of the NS mass distribution mainly relies on radio observations of Galactic NSs \citep{Ozel2016}, aided by the small number of NS events detected by LIGO/Virgo during the first three observing runs. In our work, we consider the following observationally-motivated distributions for the NS masses: 
\begin{enumerate}
\item \textit{Uniform distribution.} We consider a case where the masses of neutron stars are distributed uniformly in a range $[m_{\rm min},m_{\rm max}]$. The mass probability distribution is thus given by:
\begin{equation}
    p(m)=\frac{1}{m_{\rm max}-m_{\rm min}}
\end{equation}
with $m_{\rm min}=1\, \rm M_{\sun}$ and $m_{\rm max}=2.2\, \rm M_{\sun}$. This is consistent with the six extragalactic NS masses observed through the GW signals of BNS and BNBH detected so far \citep{Landry2021, Abbott2023pop}.

\item \textit{Gaussian distribution.} We also consider the case of a Gaussian distribution given by:
\begin{equation}
    p(m)=\frac{1}{\sqrt{2 \pi}\sigma_0}\exp\bigg[-\frac{1}{2}\bigg(\frac{m-m_0}{\sigma_0}\bigg)^2\bigg]
\end{equation}
with $m_0=1.35\, \rm M_{\sun}$ and $\sigma_0=0.15\,\rm M_{\sun}$ \citep{Borhanian2022}. We consider two options for the maximum NS mass: a lower one of $m_{\rm max}=2 \,\rm M_{\sun}$ and a higher one of $m_{\rm max}=2.5 \,\rm M_{\sun}$. We also consider a worst case scenario with $m_0=1.8\, \rm M_{\sun}$ and $m_{\rm max}=2 \,\rm M_{\sun}$. The motivation for this distribution is that, among the galactic observations, double NS systems seem to be composed by objects with a mass coming from the same tight mass distribution centered around $1.35\, \rm M_{\sun}$ \citep{Ozel2012,Kiziltan2013, Ozel2016}, although more recent works tend to disfavor the single Gaussian model \citep{Farrow2019,Galaudage2021}.

\item \textit{Double Gaussian distribution.} Evidence for a bimodal mass distribution has been reported considering the full population of galactic pulsars and would be justified by different NSs formation and evolution channels \citep{Valentim2011, Antoniadis2016, Alsing2018}. We thus consider a double Gaussian distribution parametrised by:
\begin{eqnarray}
    p(m)=A\exp\bigg[-\frac{1}{2}\bigg(\frac{m-m_1}{\sigma_1}\bigg)^2 \bigg]+\nonumber\\
    +(1-A)\exp\bigg[-\frac{1}{2}\bigg(\frac{m-m_2}{\sigma_2}\bigg)^2 \bigg],
\end{eqnarray}
with $m_1=1.32 \, \rm M_{\sun} $, $\sigma_1=0.08 \, \rm M_{\sun}$, $m_2=1.88\, \rm M_{\sun}$, ${A=0.64}$ and we normalise both Gaussian terms individually by assuming $m_{\rm max}=2.38\, \rm M_{\sun}$, such that $A$ regulates the fraction of population in each component \citep{Farr2020}. 

\end{enumerate}

\section{Results}
\label{Sec:Results}

\begin{figure*}
\centering    
\includegraphics[width=1\textwidth]{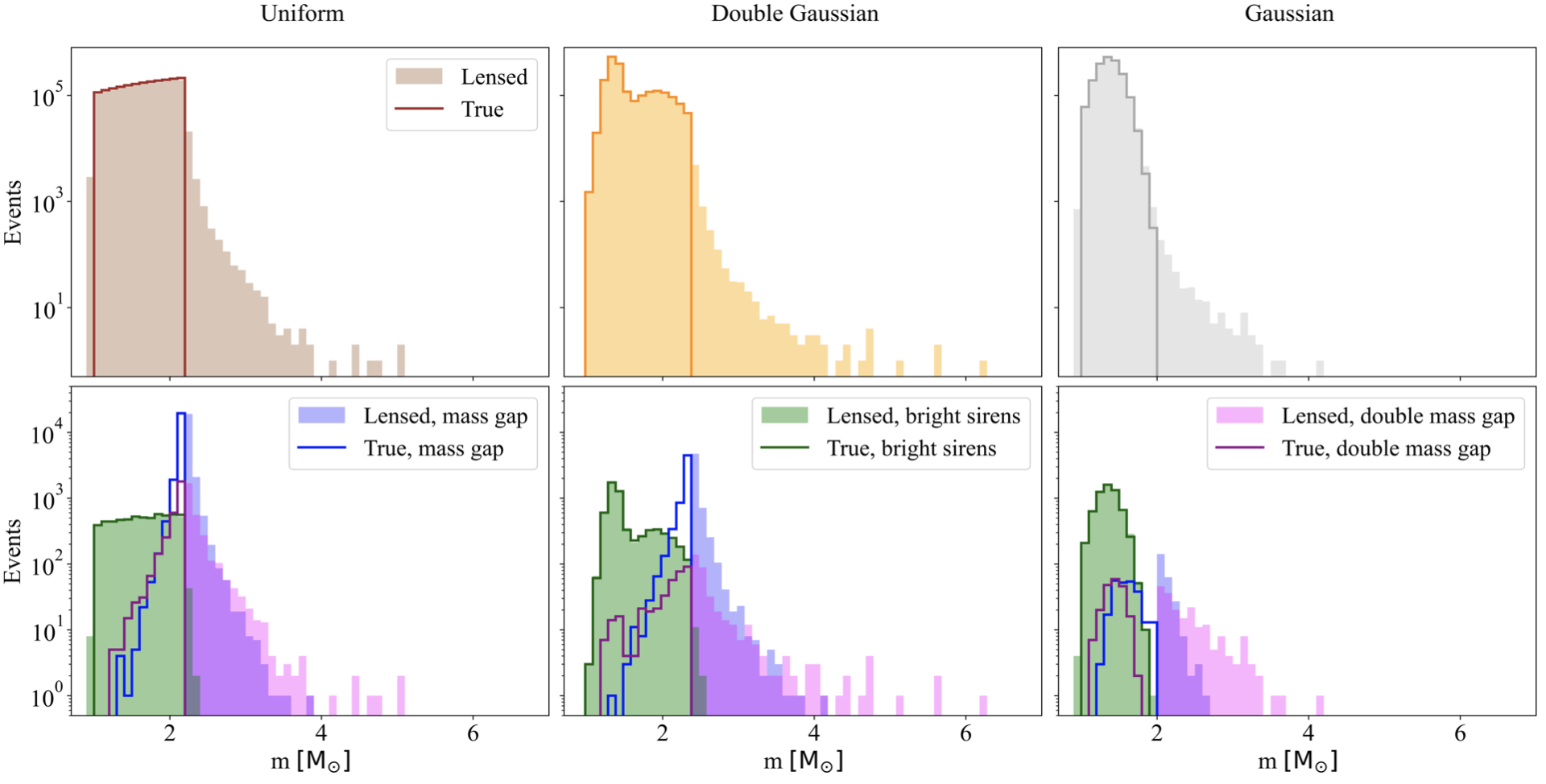}
\caption{True and lensed neutron star mass distributions, obtained simulating $10^6$ BNS events. From left to right: uniform, double Gaussian, Gaussian  with $m_{\rm max}=2\, \rm M_{\sun}$. The filled histograms represent the lensed masses, while the unfilled solid lines show the true masses. The histograms on top show the total true and lensed distributions. The histograms at the bottom show the lensed and true distributions of mass gap events and double mass gap events. The green histograms show the distribution of 3000 bright standard sirens events. Notice that in the uniform distribution case the maximum lensed mass is $8.3\, \rm M_{\sun}$ which belongs to a double mass gap event and is not shown in the plot.} 
\label{fig:results}
\end{figure*}

\begin{table}
	\centering
 \resizebox{0.48 \textwidth}{!}{
	\begin{tabular}{|l c c|} 
		\hline
		 Mass distribution &  Mass gap & Double mass gap \\    
		\hline    \hline
    	Gaussian $(m_0=1.8\,\rm M_{\sun}\; m_{\rm max}=2\,\rm M_{\sun})$ & 2835  & 458  \\
		Uniform  & 2209  & 147 \\  
        Double Gaussian & 593 & 18  \\
		Gaussian $(m_{\rm max}=2\,\rm M_{\sun})$ & 25  & 10  \\
		Gaussian $(m_{\rm max}=2.5\,\rm M_{\sun})$ & 3  & 2  \\
		\hline
	\end{tabular}}
	\caption{Number of mass gap and double mass gap events found with $10^5$ events in total, which we assume can be detected over $\sim 1 \, \rm year$ of observational period of ET, and assuming different distributions for the intrinsic NS mass. The numbers are obtained averaging over 10 realisations of the mock catalogue for the uniform, double Gaussian and Gaussian with $m_{\rm max}=2 \,\rm M_{\sun}$ and with $m_0=1.8\,\rm M_{\sun}$, over 20 realisations for the Gaussian with $m_{\rm max}=2.5\, \rm M_{\sun}$.}
 \label{tab:MGnumbers}
\end{table}

We consider the three different mass distributions defined in Subsection~\ref{subsec:mass distributions} and simulate BNS events as described in Section~\ref{Sec:Mocks}. For each mass distribution, we produce 10 realisations of $10^5$ events, except for the Gaussian case with $m_{\rm max}=2.5\, \rm M_{\sun}$, for which we consider 20 realisations. Each realisation is obtained by independently drawing values for the redshifts, masses, angles and lensing magnifications from the probability distributions described in Section~\ref{Sec:Mocks}. As said, we suppose that ET will observe $\sim 10^5$ BNS events per year. 
From the total of events, we also consider a subgroup of $300$ bright sirens per year, for which the true redshift can be estimated from the EM counterpart and used to break the mass-redshift degeneracy. As explained in Section~\ref{Sec:Mocks}, we randomly select them from the events in our mock catalogue with $z<2$ and $ \iota < 20 \degr$.

The results of our analysis are presented in Figure~\ref{fig:results}. For visual purposes, we display the histograms of $10^6$ events, obtained by summing over the 10 realisations of the mock catalogue. We show the total lensed  (shaded contours) and true (solid lines) distributions in the upper panels, and the subgroup of mass gap, double mass gap and bright sirens events in lower panels. The lensed distributions are obtained by estimating the masses of the two components of the binary system using equation~\eqref{observed masses}, while for the true distributions we break the mass-redshift degeneracy with the true redshift $z$ of the event. In both cases, we show only the observable events, i.e., the ones whose SNR is greater than $\rho_{\lim}$. 

As can be seen in Figure~\ref{fig:results}, for all three distributions, lensing creates a tail of high-mass events whose inferred masses are higher than the theoretically allowed maximum mass, and can even reach values of $\sim 5 \, \rm M_{\sun}$, which corresponds to the expectation for the least massive astrophysical BHs \citep{Bailyn1998,Farr2011, Kreidberg2012}. Furthermore, for all three distributions, we find a fraction of events whose lensed masses are below the lower mass limit, i.e., $1 \, \rm M_{\sun}$. 

In the low-mass region, the lensed masses are only slightly smaller than the true ones, with the smallest value being $0.96 \, \rm M_{\sun}$, found in the uniform distribution case. This asymmetry between the high-mass and low-mass tails of the lensed distributions is clearly consistent with the shape of the lensing magnification PDF in Figure~\ref{fig:pmu}, which is sharply peaked around $\mu=1$ and it has a tail towards higher magnifications that becomes more relevant for increasing redshifts.

In the lower panels of Figure~\ref{fig:results}, we present the lensed and true mass distributions of the mass gap and double mass gap events. As can be noticed, especially for mass gap events, the majority of events have true masses which are already close to the $m_{\rm max}$ of the distribution. In fact, massive NSs are more likely to become mass gap events, since even relatively low magnifications are enough to push the lensed masses above $m_{\rm max}$. On the contrary, the double mass events are those that have suffered higher magnification, such that the true (lensed) distributions extend to lower (higher) values of NS masses. 

\begin{figure*}
\centering    
\includegraphics[width=1\textwidth]{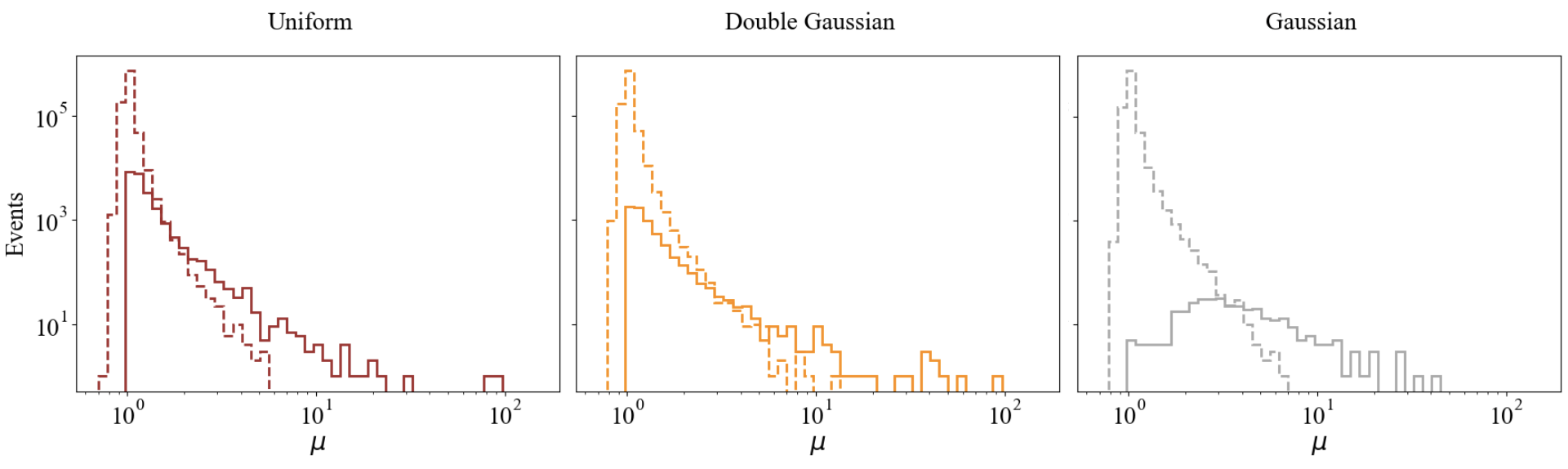}
\caption{Histograms of the lensing magnification distribution of $10^6$ events for each mass distribution. The solid lines represent the histograms of the lensing magnification of mass gap and double double mass events. The dashed lines show the histograms of the lensing magnification of the normal events, i.e. the ones whose lensed masses remain below $m_{\rm max}$. }
\label{fig:pmufull}
\end{figure*}

\begin{figure}
\centering    \includegraphics[width=0.45\textwidth]{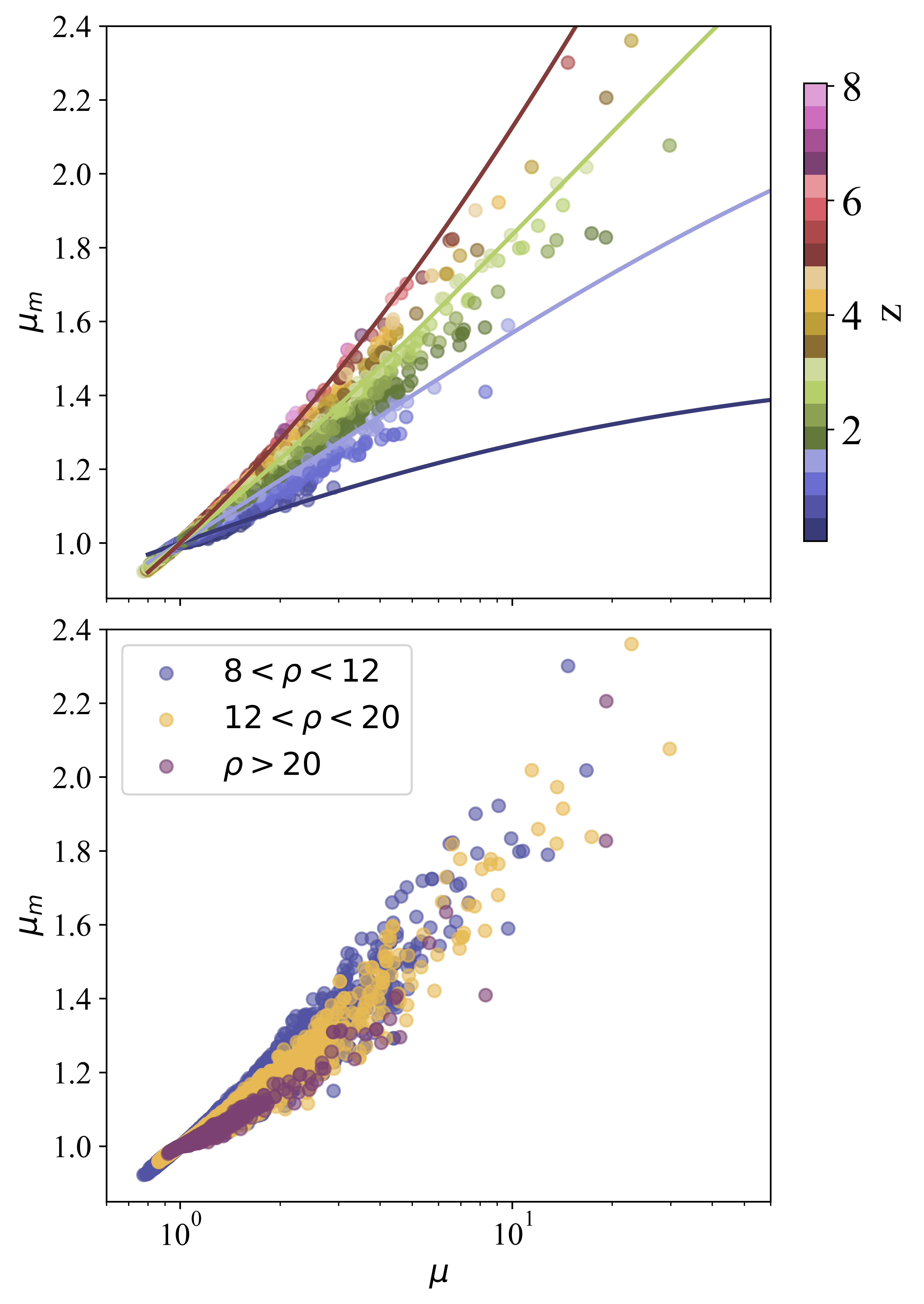}
\caption{Scatter plot of the lensing magnification $\mu$ on the x-axis and the mass bias factor $\mu_{\rm m}$ on the y-axis. In the upper panel, the colors represent the true redshift of the source. The points are obtained considering the uniform distribution for the NS masses, while the solid lines are independent on the mass distributions and are obtained numerically inverting the distance-redshift relation in equation~\eqref{dl-z} and using equation~\eqref{observed masses}. In the lower panel, the colors represent the SNR of the events.}
\label{fig:mu_m}
\end{figure}

The difference in the number of mass gap and double mass gap events between the three mass distributions can be justified by similar arguments. In Table~\ref{tab:MGnumbers}, we report the number of mass gap and double mass gap events obtained averaging over all the realisations of $10^5$ events each. We highlight that for each mass gap event we have one NS with lensed mass in the mass gap, while for each double mass gap event two NSs have lensed masses in the gap. The uniform distribution has the highest number of mass gap and double mass gap events due to the fact that it is the distribution with the highest probability near $m_{\rm max}$. Consistently, the double Gaussian and the single Gaussian follow with respectively almost one order and two orders of magnitude fewer events than the uniform one. 

We also report on the results obtained considering the Gaussian distribution with the upper value $m_{\rm max}=2.5 \, \rm M_{\sun}$ (not shown in the figures). In this case, even though the lensed distribution resembles the one obtained in the Gaussian case with $m_{\rm max}=2 \, \rm M_{\sun}$, compared to the latter case only a small fraction of events is labelled as mass gap events, due to the mismatch between the highest mass in the true distribution and the value of $m_{\rm max}$. Given the small number of mass gap and double mass gap events in the Gaussian case with $m_{\rm max}=2.5 \, \rm M_{\sun}$, we average over 20 independent realisations of the mock catalogue and we find that the scatter of the number of mass gap and double mass gap events ranges from a minimum of 1 to a maximum of 9  events.

For the Gaussian case, we also consider a worst case scenario by shifting the peak to $m_0=1.8 M_{\odot}$ while keeping $m_{\rm max}=2.0 M_{\odot}$. With these assumptions, we obtain the highest number of (double) mass gap events, and we bound the number of expected overestimated masses. 

Finally, in the lower panels of Figure~\ref{fig:results}, we show the true and lensed mass distribution of bright standard sirens, which we assume to be $300$ for each realisation of the mock catalogue. For this subgroup of events, both the true and lensed distributions can be estimated from a real catalogue of multi-messenger events, making bright sirens good candidates to perform sanity checks of the results obtained from the full catalogue. In addition, we notice that, for bright standard sirens, even the lensed mass distribution remains informative, since mass gap events are rarer and have smaller mass bias factors compared to the full catalogue. This is due to the fact that we limit the redshift of bright sirens events to be smaller than $2$. At low redshifts, high magnification values are less likely to occur and consequently also high values of the mass bias factors. 
Furthermore, by restricting $ \iota < 20 \degr$ we are selecting on-axis events which are louder in SNR, and thus less affected by selection effects, both on the mass and magnification values, which would tend to preferentially select massive events with high magnification. This is interesting for example in the case of bright sirens events with large uncertainty on the true redshift, for which the lensed mass distributions can still work as a possible check of the results obtained from the full catalogue.

In Figure~\ref{fig:pmufull}, we display the lensing magnification distributions of the events in our mock catalogue. We distinguish between the magnifications of normal events, i.e. the ones whose lensed masses remain below $m_{\rm max}$, and mass gap plus double mass gap events. As can be seen, while the lensing magnification distributions of normal events is peaked around $\mu=1$, the distributions of mass gap and double mass gap events are broader and shifted towards higher magnifications. In particular, in the limit of the Gaussian mass distribution, we notice that the magnification distribution of mass gap and double mass gap events no longer presents a peak at $\mu=1$, since these types of events have a bias towards higher magnifications.

As can be seen in Figure~\ref{fig:pmufull}, a handful of events have pretty high magnifications, i.e. $10 \lesssim \mu \lesssim 10^2$. We highlight that, in the strong lensing regime, the magnification PDF provided by \citet{Takahashi2011} is more sensitive to resolution effects of the simulations. Additionally, the approximation in equation \eqref{mu-kappa}, used to obtain the magnification PDF at $z<1$, becomes less accurate. It is then clear that the results at very high magnification regime should be interpreted with care.
We expect that some of these high-magnification events will be recognised by the strong lensing searches currently under development \citep{Haris2018,Dai2020,Wang2021, Janquart2021,Ezquiaga2021,Lo2023}. In particular, the few events at extremely high magnifications may also be characterised by wave optics effects \citep{Diego2018,Diego2019, Mishra2021, Kim2022}. Given that the rate of success of strong lensing searches in the ET era is still uncertain \citep{Caliskan2023}, our analysis may be viewed as a worst-case scenario in which most of them go undetected.

On the other hand, in the case of the uniform and double Gaussian distributions, the magnification distributions of mass gap and double mass gap events have peaks around $\mu=1$, within the weak lensing regime. While these events are expected to have lensed masses only slightly biased, they are very unlikely to be recognised by strong lensing searches, thus jeopardizing the possibility of accurately inferring the value of $m_{\rm max}$ from the data.

Lastly, we present in Figure~\ref{fig:mu_m} the relation between the lensing magnification $\mu$ and the mass bias factor $\mu_{\rm m}$. The points in Figure~\ref{fig:mu_m} correspond to the $10^6$ events whose mass distributions are given in Figure~\ref{fig:results}. While the points are generated assuming the uniform mass distribution, the solid lines in the upper panel are independent of the mass distribution and are obtained by numerically inverting the distance-redshift relation, given in equation~\eqref{dl-z}. As expected, we find a positive correlation between $\mu$ and $\mu_{\rm m}$ which depends on the redshift of the source. If we fix the value of $\mu$, we notice that at higher redshifts, higher values of $\mu_{\rm m}$ are preferred, but always with $\mu_{\rm m}<\mu$ for the redshift range considered in this work. Moreover, it is worth highlighting that in the limit $\mu \rightarrow \infty$, we have $z^{\rm biased} \rightarrow 0$, and we find the maximum of the mass bias factor to be $\mu_{\rm m}^{\rm max}=1+z$, which is the asymptote of the solid lines in Figure~\ref{fig:mu_m}.

We investigate the dependence of our results on the SNR threshold, $\rho_{\rm lim}$, set for the detection. In the uniform distribution case, we increase the threshold to $\rho_{\rm lim}=12$ \citep{Branchesi2023} and we find that the percentage of mass gap and double mass gap events reduces from $2.4\%$, obtained with $\rho_{\rm lim}=8$, to $1.8\%$.
Two opposite effects combine to yield the difference in the ratio of mass gap events to total events. With increasing $\rho_{\rm lim}$, we expect the lensing selection effect, i.e. the fact that magnified events are preferentially selected by SNR filtering over the demagnified ones, to worsen. As a consequence, increasing $\rho_{\rm lim}$, the average magnification of the events in the catalogue becomes slightly bigger. On the other hand, increasing $\rho_{\rm lim}$, we select events which are close by. This can also be noticed from the lower panel of Figure~\ref{fig:mu_m}, where we see that the points in top part of the distribution, which correspond to the high-redshift ones (see upper panel of Figure~\ref{fig:mu_m}), have lower SNR. At lower redshifts the shape of the lensing magnification PDF, as in Figure~\ref{fig:pmu}, is more sharply peaked around $\mu=1$, meaning that the scatter in the high-magnification regime becomes smaller. Moreover, at lower redshifts, lower values of $\mu_{\rm m}$ correspond to the same $\mu$. Overall, the effect of selecting events at lower redshift is stronger than the lensing selection effect. 

\section{Discussion}
\label{Sec:Discussion}

The impact of lensing on the observed neutron star mass function heavily depends on the distribution of the magnification. Therefore, it is of key importance to make some remarks on this delicate matter. Multiple sets and analytical models of the lensing magnification PDF are available in the literature \citep[see][as some examples]{Das2006, Boyle2021,Fosalba2014,Xavier2016,Cusin2019,Gouin2019,Osato2021}, but ultimately we decided to implement the simulated PDFs provided by \citet{Takahashi2011}. While higher-resolution simulations exist, they might not cover the range of redshifts required for our set-up. Even in the case of \citet{Takahashi2011}, we needed to expand the redshift range of the lensing PDF predictions using the lognormal model for the lensing convergence field provided by \citet{Marra2013} at $z<1$. While the two PDFs show excellent overall agreement at $z = 1$, there is a small discrepancy between the two at magnifications beyond $10$. 

We highlight that the strong lensing probability, given by the high-magnification tail of the distribution, is particularly sensitive to the resolution of the simulations, as can be seen from \citet[Figure 14]{Takahashi2011}. Additionally, at high magnifications, microlenses can create distortions in the magnification PDF. This effect is not modelled in our work, but it is expected to be more relevant at magnification around few hundreds, depending on the lenses configuration \citep{Diego2019b}. Moreover, we notice that multiple images of strong lensing are treated separately in our work, i.e. we assume that images of the same event will be resolved temporally and recorded as separate events. In fact, the magnification PDF that we use is obtained from the sum of the PDFs of the different type of images. On the other hand, we do not account for the fact that strongly lensed events have images with correlated magnifications. We leave for future work a more accurate modelling of the strong lensing cases, which can be treated similarly to strongly lensed supernovae \citep{Goldstein2017, Arendse2023,Suyu2024}. 

A second point to consider is the fact that an interpolation had to be used for intermediate redshift values.  While this interpolation might not precisely reflect the true distributions it was a compromise that had to be made, as making a distribution for every redshift value is computationally unfeasible. Finally, we highlight that baryonic effects are not included in the simulations we consider and they may impact the magnification distribution, especially its tail. Including baryons, \citet{Osato2021} finds a $\sim 10 \%$ difference in the convergence field distribution, with a smaller impact at higher redshift. With these points in mind, our results might slightly under- or overestimate the number of (double) mass gap events, and their relative ratio, we would expect for the future ET's measurements, due to the real lensing magnification PDF differing from the one used, but we expect the overall conclusions to be robust.

Our results show that ET is expected to detect some events for which the lensed masses exceed $m_{\rm max}$. Considering the uniform distribution, we can check that the majority of mass gap events are at high redshifts: among all the events with a lensed mass higher than $m_{\rm max}$ we find that $\sim 1.4\%$ have $z<0.5$, $\sim 36.4\%$ have $0.5<z<1.5$ and $\sim 62.1\%$ have $z>1.5$. Previous works, such as \citet{Broadhurst2020} and \citet{Bianconi2023}, investigated the hypothesis that some of the recent LIGO/Virgo observations with NS masses compatible to fall within the mass gap, might be unidentified lensed events. With our work, we can neither confirm nor deny whether the estimation of the masses of some special events, such as GW190814, have been significantly impacted by lensing. Similarly, we cannot make predictions on mass gap events for the ongoing and upcoming runs of LIGO/Virgo \citep[for a discussion on this see][instead]{Graham2023}. The fifth observing run O5 is expected to start before the end of this decade and, at final target sensitivity, might see a few hundreds of BNS events per year up to $z \simeq 0.2$ \citep{Abbott2020c,Kiendrebeogo2023}. At these redshifts, the PDF of the lensing magnification is very sharply peaked around $\mu=1$ such that for the expected number of events our work suggests that the lensing hypothesis is very weak. Additional work, specifically focused on modelling the strong lensing probability, is needed to better address this point.  

What we can confirm is that, much like the conclusion drawn by \citet{Dai2017}, \citet{Oguri2018} and \citet{He2022} on the black hole mass function, the high-mass end of the neutron star mass distribution is influenced by lensing. This effect will be measurable for ET and other 3G detectors. However, unlike for black holes, the maximum mass of the NSs is more well-defined and has big implications on the EoS. Moreover, lensed BNS might be misinterpreted as binary BH events due to the high observed masses. Since only BNS and NSBH mergers are expected to be accompanied by EM emission, particular care should be used when selecting the mass criteria for sending alerts for multi-messenger searches. We highlight that our results are obtained considering a source by source analysis. Other statistical approaches such as in \citet{Congedo2019} may be used to constrain the parameters of the mass distribution of a population of NSs.

Efforts to develop methods to infer the true redshift of dark sirens events have been pursued mainly in view of studying cosmology and possible extensions of its standard model \citep[see][and references therein]{DelPozzo2012,Gray2020,Mukherjee2022,Palmese2023, Gray2023,Ghosh2023}. These methods could break the mass-redshift degeneracy without using information from the lensed luminosity distance. In particular, the ``spectral siren method'' relies on the comparison of the source-frame and the detector-frame distribution of parameters of the GW populations, especially the mass distribution \citep{Taylor2012, Ezquiaga2022}. To apply this method, it may become important to investigate how to mitigate the impact of lensing on the mass distribution.

Finally, it is worth highlighting that our results are obtained considering the cosmology and the redshift distribution in equation \eqref{pth} as fixed and perfectly known. Mismodelling the redshift distribution can change the fraction of (double) mass gap events. Additionally, future works may investigate the dependence of our results on the cosmology, by accounting for the uncertainty on the cosmological parameters, especially the Hubble constant, when inverting the distance-redshift relation. 

\section{Conclusions}
\label{Sec:Conclusion}

Future prospects of gravitational wave detections are excellent for third generation GW observatories like ET. Due to the high increase in sensitivity compared to currently operational detectors, ET is expected to detect more numerous as well as more distant BNS mergers. However, this also implies that the number of significantly lensed events (i.e., those whose amplitude has been altered by fluctuations of the matter distribution in the universe) is also expected to grow drastically. 

In this work, we quantitatively investigated the impact of (weak and strong) lensing on the NS mass distribution observed through GWs from BNS mergers. While we find that the overall shape of the mass distribution is preserved, lensing induces a high-mass tail in the distribution. This shows that if lensed events are not identified as such, the observed maximum mass ascribed to neutron stars would be biased. This in turn could affect our understanding of nuclear matter and its equation of state in the interior of neutron stars. Therefore, it is crucial to understand the extent to which the observed NS masses differ from their true values.

To perform our analysis, we generated $10$ mock catalogues consisting of $10^5$ BNS merger events each, for several possible intrinsic NS mass distributions. For each event, we amplified the signal by a factor of $\sqrt{\mu}$, where $\mu$ is the redshift-dependent lensing magnification drawn from simulated PDFs. We broke the mass-redshift degeneracy using the distance obtained from the lensed amplitude of the signal, we derived the observed neutron star masses and compared them to the truth. 

Our results show that a fraction of the events has at least one component with observed mass above the theoretical maximum mass $m_{\rm max}$, and that this fraction depends greatly on the intrinsic mass distribution. The scenario that is most biased by lensing assumes a uniform mass distribution, and have on average $2.2\%$ of events being mass gap events, and $0.15\%$ double mass gap events. The least impact is found for the catalogues with a Gaussian true mass distribution ($m_{\rm max} = 2.5\, \rm M_\sun$), which show only a handful of mass gap events. Events whose observed mass are above the maximum are normally already near the maximum mass of the intrinsic distribution. In particular, we found that mass gap events tend to have one of the components with the true mass already close to the maximum, such that they are more likely to be only slightly magnified, whereas in comparison, double mass gap events tend to be more highly magnified events with lower intrinsic masses.

In each catalogue, we considered a subset of $300$ bright siren events, for which the true redshifts are known, such that we know both the true and lensed masses. For these events the lensed mass distribution and the true mass distribution shows a greater resemblance than for dark siren events. This is mainly due to the fact that for an EM counterpart to be observed, the source must be relatively close by (we assumed $z<2$), in which case high magnifications are overall less likely.

In conclusion, we showed that ET is expected to detect anywhere from $5$ to $2500$ BNS mergers per year where at least one of the component masses appears to be higher than the theoretical maximum for the mass of a neutron star. Based on these results, we consider it will be crucial to take this into account when investigating nuclear physics and cosmology through GWs.

\section*{Acknowledgements}

We thank Tanja Hinderer for helpful comments throughout our research process. 

\section*{Data Availability}

The files containing the mock catalogue of events used in this paper are publicly available at the following link: \href{https://github.com/Canev001/GWlensing_NSmasses}{Github/GWlensingNSmasses}.


\bibliographystyle{mnras}
\bibliography{main}

\end{document}